\documentclass[
  journal=pasa,
  manuscript=research-article,
  year=2023,
  volume=xx,
]{cup-journal}

\usepackage{amsmath}
\usepackage[nopatch]{microtype}
\usepackage{booktabs}
\usepackage{hyperref}
\usepackage{subcaption}

\title{RFI Detection with Spiking Neural Networks}

\author{N. J. Pritchard}
\affiliation{International Centre for Radio Astronomy Research, University of Western Australia, Perth, 6009, Western Australia, Australia}
\email[N. J. Pritchard]{nicholas.pritchard@icrar.org}

\author{A. Wicenec}
\affiliation{International Centre for Radio Astronomy Research, University of Western Australia, Perth, 6009, Western Australia, Australia}

\author{M. Bennamoun}
\affiliation{School of Physics, Mathematics and Computing, University of Western Australia, Perth, 6009, Western Australia, Australia}

\author{R. Dodson}
\affiliation{International Centre for Radio Astronomy Research, University of Western Australia, Perth, 6009, Western Australia, Australia}


\begin{document}

\begin{abstract}
Detecting and mitigating Radio Frequency Interference (RFI) is critical for enabling and maximising the scientific output of radio telescopes.
The emergence of machine learning methods capable of handling large datasets has led to their application in radio astronomy, particularly in RFI detection.
Spiking Neural Networks (SNNs), inspired by biological systems, are well-suited for processing spatio-temporal data.
This study introduces the first exploratory application of SNNs to an astronomical data-processing task, specifically RFI detection.
We adapt the nearest-latent-neighbours (NLN) algorithm and auto-encoder architecture proposed by previous authors to SNN execution by direct ANN2SNN conversion, enabling simplified downstream RFI detection by sampling the naturally varying latent space from the internal spiking neurons.
Our subsequent evaluation aims to determine whether SNNs are viable for future RFI detection schemes.
We evaluate detection performance with the simulated HERA telescope and hand-labelled LOFAR observation dataset the original authors provided.
We additionally evaluate detection performance with a new MeerKAT-inspired simulation dataset that provides a technical challenge for machine-learnt RFI detection methods.
This dataset focuses on satellite-based RFI, an increasingly important class of RFI and is an additional contribution.
Our SNN approach remains competitive with the original NLN algorithm and AOFlagger in AUROC, AUPRC and F1 scores for the HERA dataset but exhibits difficulty in the LOFAR and Tabascal datasets. 
However, our method maintains this accuracy while completely removing the compute and memory-intense latent sampling step found in NLN.
This work demonstrates the viability of SNNs as a promising avenue for machine-learning-based RFI detection in radio telescopes by establishing a minimal performance baseline on traditional and nascent satellite-based RFI sources and is the first work to our knowledge to apply SNNs in astronomy.
\end{abstract}
\section{Introduction}
Radio Frequency Interference (RFI) poses significant challenges for current and future radio telescopes, with RFI sources' number, variety, and overall disruption to observations increasing (\cite{noauthor_report_2022}). Consequently, detecting and mitigating RFI is critical for modern radio observatories. RFI detection algorithms can, broadly speaking, be categorised as either analytical or machine-learning-based. Although machine-learning methods have shown promise in outperforming traditional approaches, they often require vast amounts of labelled training data, obtained either through expensive manual annotation or from traditional algorithms, as \cite{akeret_radio_2017} do.
For these reasons and their relative computational expense, the practical deployment of machine-learning-based RFI detection schemes remains limited, necessitating fine-tuning algorithm parameters and manual data annotations for optimal performance. However, the ability to learn from the abundance of available interferometric data is tantalising, especially in an increasingly crowded radio background.

The Nearest-Latent-Neighbours (NLN) algorithm proposed by \cite{mesarcik_learning_2022} introduces a novel approach to RFI detection by framing it as a downstream anomaly detection task based on a generative model trained to represent only noiseless radio data as flagged by AOFlagger produced by \cite{offringa_aoflagger_2010}. This methodology significantly reduces the dependence on high-quality flags in the training data, making it more suitable for real-world deployment. Nonetheless, the NLN algorithm requires sampling similar data patches from a pre-loaded test set, requiring the availability of this data in memory during inference and ensuring it represents the operational environment.

In parallel to the development of ANNs, Spiking Neural Networks (SNNs), a computational model inspired by biological neuron behaviours, are becoming increasingly practical and affordable to simulate in standard computing hardware or implement in specialised neuromorphic processors for various machine learning tasks. \cite{bouvier_spiking_2019} survey recent developments and the foundations of SNNs.
The spiking behaviour and time-varying nature of SNNs make them particularly well-suited for spatio-temporal data processing, rendering them promising candidates for astronomical data-processing tasks. SNNs can be trained from scratch with various methods, including biologically inspired rules, adaptations of back-propagation, or evolutionary algorithms. Additionally, SNNs can be generated by converting pre-trained Artificial Neural Networks (ANNs), exploiting the advantages of ANN methodologies while leveraging the energy-efficiency and time-varying dynamics of SNNs at inference at the cost of foregoing their unique capabilities during training. \cite{yi_learning_2023} provide an excellent review of such methods.

To evaluate the suitability of SNNs for astronomical data processing, particularly RFI detection, we propose converting the NLN algorithm into the Spiking Nearest Latent Neighbours (SNLN) method. This approach involves training a generative model using the same methodology as the original NLN scheme and then converting the network into an SNN. Instead of searching for latent-neighbours from available data, SNLN averages a set of outputs from the SNN auto-encoder after presenting the network with a single input patch for multiple time steps. This process effectively samples the internal latent space, which exhibits stochastic variation over time, exploiting the spiking behaviour to generate its own latent neighbours.
Our method retains the operational benefits of the original NLN approach in terms of RFI detection compared to supervised ANN methods while also reducing the amount of data required to evaluate each inference patch to the specific patch under consideration.

This study presents four contributions to (radio) astronomy and RFI detection.
\textbf{Firstly}, we introduce Spiking Neural Networks (SNNs) to the domain of astronomy for the first time, opening up new avenues for leveraging this biologically inspired computational model in data processing tasks for astronomical data analysis.
\textbf{Secondly}, we propose a novel RFI detection algorithm, named Spiking Latent Neighbours (SNLN), which builds upon the unsupervised approach of the Nearest-Latent-Neighbours (NLN) method. We train an ANN model suitable for the NLN method, convert it to an SNN and modify the downstream inference task to use SNNs' time-varying nature.
\textbf{Thirdly}, we introduce a new RFI detection dataset based on the MeerKAT telescope produced by the Tabascal simulation package written by \cite{finlay_trajectory-based_2023} focussing on satellite-based RFI.
\textbf{Finally}, we evaluate the effectiveness of our SNLN approach compared to the original Artificial Neural Network (ANN) method, providing valuable insights into the detection performance trade-offs and highlighting the competitive nature of our SNN-based approach.
Collectively, these contributions showcase the promise of SNNs in improving RFI detection in radio telescopes and underscore their potential as a valuable tool for future astronomical data processing.

Section \ref{sec:related} briefly discusses other RFI detection schemes, more thoroughly introduces SNNs, discusses general anomaly detection using SNNs and discusses auto-encoder architectures with SNNs. Section \ref{sec:methods} explains the ANN2SNN conversion and novel SNLN algorithm in more detail. Section \ref{sec:dataselection} discusses the data used in training and evaluation and introduces our new Tabascal dataset. Section \ref{sec:results} presents the results of our experiments, and we present our conclusions and several suggestions for future work in section \ref{sec:conclusion}.

\section{Related Work}\label{sec:related}
This section discusses the application of neural networks to RFI detection. Astronomy is a good target for deep learning techniques due to the abundance of available data. Specifically, neural networks have been used for galaxy detection, classification, and transient detection. RFI detection can be viewed as a classification task on pixels in time-frequency spectrograms, aiming to output a boolean mask of contaminated pixels, the same as traditional algorithms output. This task, more generally known as semantic segmentation, is also applied to other domains like medical imaging.
\subsection{RFI Detection with Neural Networks}
The most popular base architecture for supervised RFI detection with neural networks is UNet, a neural network designed for image segmentation. Most works modify the original network design, improvements to the training routine, or both such as \cite{yang_deep_2020}. As is the case in all ML works, data is central to overall performance. Many works use simulated data for training, providing high confidence in the labels but limiting the model to the constraints of the simulator's model.
On the other hand, works using real data face a choice of using a traditional algorithm to generate training flags or using expertly labelled data. Using traditional algorithms to generate training masks on simulated or real data is straightforward. It yields abundant training data but constrains the ANN to learn from the behaviour of the traditional algorithm. Using expertly labelled data, arguably the most accurate by definition, is expensive, scarce, and, therefore, strains the training scheme. An alternate strategy is to train on simulated data and then perform transfer learning with small amounts of expertly labelled real data, aiming to combine the benefits of all worlds (e.g., \cite{vafaeisadr_deep_2020}) often termed Sim2Real. This Sim2Real method often struggles with a reality gap between the simulation behaviours and the real world, adding an additional challenge.

\cite{mesarcik_learning_2022} introduce a particularly interesting approach, where a traditional flagging algorithm generates flags with many false positives to select uncontaminated patches with high certainty, training an auto-encoder with these uncontaminated patches, then using this model to perform RFI detection as a downstream inference task, transforming the task from a semantic segmentation problem into an anomaly detection problem. By adopting this method, an abundance of training data is available, agnostic to the noise sources in the real data. This approach serves as the basis for our SNN conversion. However, the anomaly detection routine, which involves searching through latent representations of other examples, is iterated upon with our SNN approach while retaining the appealing innovations introduced by this method.

Developing a general-purpose model capable of producing RFI masks for any telescope across a wide range of frequencies and configurations remains a grand challenge in applying ML and NNs to RFI detection. Part of the enduring appeal of using traditional algorithmic approaches is their flexibility. Despite extensive efforts, currently, no method, traditional or NN-based, achieves an F1-score above $0.6$ on a real, representative dataset \cite{mesarcik_learning_2022}. As we explore and develop new approaches, achieving better performance on real-world data becomes a crucial focus in advancing RFI detection techniques using machine learning and neural networks in radio astronomy.
\subsection{Spiking Neural Networks}
Spiking neural networks (SNNs) are artificial neural networks that closely mimic the behaviour of biological neurons, enabling efficient and accurate processing of complex data sets. Unlike traditional artificial neural networks, which use continuous activation functions, SNNs use discrete events called spikes. Incoming spikes contribute to a neuron's membrane potential, and when the potential reaches a threshold, the neuron emits a spike. The precise mathematical behaviours of how the membrane potential increases with spike arrivals and decays over time hold much of the basis for SNNs and contemporary ANNs.

The leaky-integrate-and-fire (LiF) neuron is a simple yet accurate model of this biological process, resembling a low-pass filter circuit with a resistor $R$ and capacitor $C$. The following $RC$ circuit equation models the dynamics of the passive membrane potential for a single neuron subject to an input current $I$:
\begin{equation}
    \tau\frac{dU(t)}{dt} = -U(t) + I_{in}(t)R,
\end{equation}
where $\tau = RC$ is a fixed time constant. \cite{laurence_f_abbott_theoretical_2001} provide a comprehensive derivation of spiking neuron dynamics and their relationship to artificial neural networks.

SNNs' close resemblance to biological neurons allows them to excel at tasks challenging for other neural network types as \cite{deng_rethinking_2020} show, particularly real-time spatio-temporal data processing.
It is precisely the time-varying nature of spiking neurons that contributes to their ability to capture time-varying details of ingested data.
This spiking behaviour enables various information encoding techniques based on the timings of individual spikes or patterns of spikes, unlike artificial neurons in most contemporary deep learning approaches that operate only on an equivalent to rate-based encoding.

It is worth noting that these two neural network models are equivalent, with artificial neural networks better suited for implementation on existing computing hardware. The time-varying nature of SNNs requires specialised hardware to realise efficiently but consumes orders of magnitude less energy in the process, \cite{schuman_opportunities_2022} give a wide and long-range view of a future for neuromorphic computing. The equivalence rooted in the derivation of sigmoidal neurons found in ANNs from spiking neurons permits the conversion of trained ANNs back to an SNN as \cite{Rueckauer2016TheoryAT} summarise.

The development of neuromorphic hardware and spiking neural networks has occurred alongside contemporary deep learning and artificial neural network research. \cite{james_historical_2017} summarises the main developments throughout the last century. The increasing importance energy efficiency plays in modern computing reignites more widespread interest in SNNs. While energy improvements without rigorous benchmarking on real hardware are difficult, \cite{lemaire_analytical_2023} provide an analytical approach to estimate the energy use of equivalently sized ANNs and SNNs and show an improvement of six to eight times greater efficiency for SNNs on 45nm CMOS hardware.

The advancement in neuromorphic hardware and SNNs holds promise for tackling energy-intensive tasks while preserving the advantages of spiking behaviours inspired by biological neurons.
Neuromorphic hardware combined with a suitable SNN-based technique would enable real-time and adaptable RFI detection, with orders of magnitude less power consumption than contemporary ANN and classical methods.
\subsection{Anomaly Detection with SNNs}
Anomaly detection is a problem often encountered in time-varying domains. As such, there are several works exploring anomaly detection with SNNs. Notably, many anomaly detection works employing SNNs exist in sensor-based environments and on one-dimensional time-varying data streams.
In healthcare, \cite{bauer_real-time_2019} employ a recurrent SNN and custom neuromorphic processor to preprocess electrocardiograms for detecting anomalous behaviour. In industrial settings, \cite{demertzis_spiking_2017} implement one-class anomaly detection with SNNs for industrial control systems, while \cite{demertzis_gryphon_2020} propose a semi-supervised one-class anomaly detection scheme using SNNs in industrial environments. \cite{dennler_online_2021} design neuromorphic hardware for real-time detection of anomalous vibrations, applicable to various industrial settings.

The conversion of Artificial Neural Networks (ANNs) to SNNs for anomaly detection is also explored. \cite{jaoudi_conversion_2020} convert a traditional auto-encoder to an SNN to detect in-vehicle cyberattacks, achieving comparable performance to the original converted ANN in analysing one-dimensional, time-varying data. \cite{paul_energy-efficient_2022} convert a one-dimensional convolutional neural network to an SNN for anomaly detection in the breathing patterns of newborn infants.

Several studies present hardware implementations for real-time anomaly detection using SNNs. \cite{chen_real-time_2017} propose a component library and design method for implementing real-time anomaly detection on IBM's TrueNorth neuromorphic chip. \cite{ronchini_cmos-based_2022} develop wearable neuromorphic hardware for detecting anomalies in biological measurements in healthcare settings.

Other notable works include \cite{chen_anrad_2018} presenting an anomaly recognition and detection framework for incremental online learning from data streams, \cite{stratton_spiking_2020} performing anomaly detection on text streams with SNNs, \cite{maciag_unsupervised_2021} providing an online evolving SNN method for detecting anomalies in streaming data without supervision, \cite{phusalculkajom_multiple_2022} using SNNs to generate fuzzy logic intervals for detecting anomalies in rail networks, and \cite{baessler_unsupervised_2022} proposing a multivariate anomaly detection scheme based on SNNs.

Our approach uniquely sits as a conversion of an existing auto-encoder performing anomaly detection on two-dimensional spectrograms.
\subsection{Auto-Encoders with SNNs}
Auto-encoders are a neural network architecture designed to extract signals from high-dimensional feature space without supervision. The encoder compresses raw input data into a latent space, and the decoder reconstructs the original input data from this latent representation. Training an auto-encoder minimises the difference between the reconstruction and input data, enabling it to learn an efficient representation of the input data.

Efforts have been made to craft SNN-based auto-encoders to achieve energy-efficient and sparse latent representations. For instance, \cite{alom_network_2017} trains an auto-encoder for detecting anomalous network packets, converting it to an SNN using discrete vector factorisation, which is then simulated and executed on IBM TrueNorth hardware. \cite{yin_non-negative_2019} combine an auto-encoder architecture with a spiking random neural network. Their approach performs well on typical image datasets. \cite{roy_synthesizing_2019} develop a spiking auto-encoder to synthesise audio inputs into image outputs. \cite{stratton_spiking_2020} perform anomaly detection on text streams using an SNN auto-encoder. \cite{kamata_fully_2022} present a fully spiking variational auto-encoder, enabling variational learning through random sampling of the latent space. \cite{stewart_encoding_2022} create a hybrid SNN auto-encoder specialised in encoding event-based data from vision sensors.

In our work, we leverage the time-varying nature of the latent state built by an SNN during inference to simplify the downstream anomaly detection scheme.
\section{Methods}\label{sec:methods}
In Summary, we train an auto-encoder in the same way as applicable for the NLN algorithm, convert this trained ANN to an SNN and modify the downstream inference task. The modified inference task is more computationally efficient than the original NLN owing to the unique time-varying property of SNNs.
We adopt the same problem formulation for fair and reasonable comparison as \cite{mesarcik_learning_2022}. We re-implemented the model, training loop, and evaluation routines in PyTorch, replicating the original study. To promote transparency and accessibility, we make our code, evaluation datasets, and all data used in the figures presented in this paper available online.
Our analysis aims to provide a minimal detection performance baseline against which future methods can be compared. This decision guides us in selecting datasets and tests for which performance metrics are known in the case of the HERA and LOFAR datasets or providing a simple yet challenging test in the case of our Tabascal dataset. We do not assess computational performance in this work as we aim to focus on exploring the viability of SNNs for RFI detection.  

\subsection{Conversion to SNN}
Several SNN simulation and training libraries exist, and we selected SpikingJelly by \cite{fang_spikingjelly_2020} due to its robust GPU acceleration and straightforward ANN-to-SNN (ANN2SNN) conversion routine. The conversion algorithm replaces neuron layers with spiking neurons and scales existing weights to voltage values suitable for the LiF neuron implementation. Successful conversion requires using ReLU activation functions and the absence of MaxPool layers, both of which are true in the original network architecture.

\subsection{Network Architecture}
Our network architecture is a two-layer strided auto-encoder, as in the original work. We include optional batch-norm and dropout layers after each convolutional layer for regularisation. Regularisation is an additional parameter. 

\subsection{Training Strategy \& Hyper-Parameter Optimisation}
\begin{table*}[htbp]
\centering
\caption{Results of hyper-parameter search. The number of trials conducted for each dataset is listed in parentheses next to the dataset name. LOFAR optimisation is limited due to extensive training length.}
\label{tab:hyperparameters}
\begin{tabular}{ccccc}
\hline
Attribute          &  Hera (256)& LOFAR (32)&Tabascal (256)&Parameter Range\\ \hline
Latent Dimension            & 64& 64 & 64 &8, 16, 32, 64\\ \midrule
Batch Size                  & 32& 128 & 32 &16, 32, 64,  128\\ \midrule
Epochs                      & 17& 100 & 75 &2, 128\\ \midrule
Auto-Encoding Learning Rate & 1.59e-3& 1e-4 & 2.00e-4 & 1e-4 - 1e-2\\ \midrule 
Generator Learning Rate     & 1.77e-3& 1e-4 & 7.76e-4 &1e-4 - 1e-2\\ \midrule
Discriminator Learning Rate & 1.68e-4& 1e-4 & 6.87e-4 &1e-4 - 1e-2\\ \midrule
Optimizer                   & Adam & Adam & Adam &Adam, RMSprop, SGD\\ \midrule
Neighbours                  & 21& 20 & 20 &1 - 25\\ \midrule
Num Filters                 & 16& 32 & 64 &16, 32, 64\\ \midrule
Regularize Network& False& True & True &True, False\\
\end{tabular}%
\end{table*}
We diverge from the original training strategy \cite{mesarcik_learning_2022} deploy by performing a hyper-parameter search individually for each dataset. 
We used Optuna written by \cite{akiba_optuna_2019} for hyper-parameter optimisation; we specifically use the default tree-structured Parzan estimation algorithm to intelligently search through potential options. The hyper-parameter optimisation generates hyper-parameters comparable to the original work with adjustments to the number of training epochs. We optimise hyper-parameters to yield the highest F1-score; we look for the best-performing auto-encoder, end to end to ensure the network is informative for latent neighbour searching.
Table \ref{tab:hyperparameters} displays our final hyper-parameters.
The requirement for larger learning rates and epochs for the LOFAR dataset reflects the increased challenge of detecting RFI in real data.
Determining that the number of neighbours is almost identical may suggest that the size of patches sent through the auto-encoder determines this attribute.

\subsection{Spiking NLN}
We encourage readers to refer to \cite{mesarcik_learning_2022} for an authoritative explanation. However, the original nearest-latent-neighbours (NLN) algorithm operates as follows:
\begin{itemize}
    \item A set of test patches are encoded into latent space.
    \item Additionally, the set of training patches is additionally encoded into latent space.
    \item For each latent-test-patch, an L2 norm search is performed against the latent-training-patches to find the $k$ most similar.
	\item The test patches and training patches are decoded.
	\item The $k$ most similar test patches are averaged and subtracted from the test.
	\item The remaining pixel values in the test patches are binarised to obtain a noise mask.
\end{itemize}

In contrast, our Spiking Nearest Latent Neighbours (SNLN) algorithm utilises the time-varying nature of SNN execution to build implicitly similar latent neighbours. SNLN operates by:
\begin{itemize}
    \item Presenting each test patch to the SNN auto-encoder for $T$ timesteps
    \item The final $k$ output frames from the auto-encoder represent the decoded values of the $k$ internal latent representations.
    \item Average the $k$ output frames and subtract the result from the original test patch, and the remaining pixel values are binarised to obtain a noise mask.
\end{itemize}
The SNLN method is notably simpler than the original NLN as it relies solely on the test data of interest at inference time. This is possible due to the time-varying nature of SNN execution.
By skipping the expensive latent-neighbour search found in the original NLN, the SNLN requires a constant, rather than variable, quantity of computing resources regardless of the quantity flagged data, addressing the main weakness known of the original NLN.  
\section{Data Selection}\label{sec:dataselection}
Existing machine learning (ML) approaches to Radio Frequency Interference (RFI) detection typically rely on abundant high-quality labelled data. By transforming the RFI detection problem from a semantic segmentation problem to an anomaly detection problem \cite{mesarcik_learning_2022} significantly reduces the need for large-volumes of high-quality labelled data. Additionally, inconsistency with test data is a common issue in comparing RFI detection methods. To ensure a fair comparison, we utilise the same datasets prepared by \cite{mesarcik_learning_2022-1}, specifically the Hydrogen Epoch of Reionisation Array simulations \cite{deboer_hydrogen_2017} and hand-labelled LOFAR observations. The only modification required to adapt the datasets for use with PyTorch is changing the channel containing pixel values from the last channel to the first channel; otherwise, our dataset is prepared in precisely the same manner as the original work.
\subsection{Tabascal}
We additionally create and publish openly a new dataset generated using the Tabascal simulator written by \cite{finlay_trajectory-based_2023} and emulating the MeerKAT telescope.
The main focus of this dataset is to provide a large-array mid-frequency dataset that includes moving low-orbit satellite RFI sources, which are expected to become increasingly troublesome in the coming years.
The majority of simulation parameters come from \cite{finlay_trajectory-based_2023} which are in turn derived from MeerKAT documentation.
We simulate for 512, two second integration steps over 512 frequency channels from $1.227$GHz to $1.334$GHz of $209$ KHz width, average the visibility amplitude over $16$ samples per integration and finally sample 512 baselines to produce $512 \times 512 \times 512 \times 1$ datasets.
The resulting spectrograms contain RFI localised in narrow frequency bands but present for the entire simulation duration. While this dataset is not truly indicative of practical observation, and in practice, these frequencies may be removed entirely without the need for more sophisticated methods, it creates a difficult task for machine-learning methods to distinguish the fine boundary between the contaminated and uncontaminated region precisely.

\begin{table}[htbp]
\centering
\caption{Description of each dataset used for training and testing. We reproduce values for the HERA and LOFAR datasets from \cite{mesarcik_learning_2022}.}
\label{tab:datasetsizes}
\begin{tabular}{ccccc}
\hline
Dataset &  \# Baselines & \# Training samples & \# Test Samples & \% RFI \\ \hline
HERA    & 28  & 420 & 140 & 2.76 \\ \midrule
LOFAR   & 2775& 7500 & 109 & 1.26 \\ \midrule
Tabascal& 512 & 409 & 103 & 12.23\
\end{tabular}%
\end{table}

Table \ref{tab:datasetsizes} contains the size of each dataset and the proportion of RFI contamination.

Tabascal's architecture produces sky and RFI visibilities separately, which we then use to produce input and label data.
Access to machine-precision ground truth in the RFI visibilities introduces an interesting problem; the vast majority of affected pixels are perturbed by incredibly small values close to zero.
Therefore, we can threshold the ground-truth data to provide the same dataset at several difficulty levels.
Our uploaded datasets contain the original RFI visibility data, masks without thresholds, with thresholds of $0, 1, 2, 4, 8, 16$ and finally thresholded above the median value of the sky visibilities, as a rough analogue of human perception.
We use a simulation containing two satellite sources similar to GPS and GLONASS satellites and three static ground-based RFI sources.

The resulting dataset exhibits RFI strongly localised in time but over all frequencies and thus provides a challenge for the NLN method, which will never be exposed to heavily contaminated data during training.  

\section{Results}\label{sec:results}
We compare the results of our implementation of NLN and our new algorithm SNLN on simulated HERA data, hand-labelled LOFAR data and Tabascal-simulated MeerKAT data. We recreate the AOFlagger threshold and out-of-distribution (OOD) experiments found in the original work for the HERA dataset. For the LOFAR and Tabascal datasets, we report results from repeat trials containing all available data. To assess the overall performance of our algorithms on the downstream RFI detection task, we employ three common metrics: Area Under the Receiver Operating Characteristic (AUROC), Area Under Precision-Recall Curve (AUPRC), and F1-Score.

AUROC measures the model's ability to distinguish between positive and negative classes by varying the classification threshold. A higher AUROC indicates better discrimination between the two classes. However, in our case, the AUROC score is heavily influenced by the behaviour on uncontaminated pixels.

AUPRC is particularly relevant in our task, considering the imbalanced nature of RFI detection, where RFI examples are relatively infrequent compared to the rest of the data. The AUPRC score evaluates the trade-off between precision (the proportion of correctly classified positive examples out of all positively classified examples) and recall (the true positive rate). It quantifies the model's ability to identify RFI while minimising false positives correctly. Additionally, the minimum AUPRC score is proportional to the number of positive examples in a dataset; datasets with less RFI will have a lower acceptable AUPRC score, reflecting the challenges of imbalanced data distributions.

The F1-Score is the harmonic mean of precision and recall, providing a balanced measure of the model's overall performance. By considering both precision and recall, the F1-Score helps us to evaluate how well the model achieves a balance between correctly identifying RFI and minimising false positives, thereby providing a comprehensive assessment of its efficacy in RFI detection tasks.

The AOFlagger strategies used in performance comparisons are available online\footnote{https://github.com/mesarcik/RFI-NLN} and are based on standard example strategy files.
The strategy comprises a highpass filter and a transient filter, parameterised with a single threshold variable.
While more complex features are not present, we include a comparison with AOFlagger to show the adaptability of ML-based methods and do not provide a comprehensive benchmark here.

\subsection{HERA}
\begin{figure*}
    \centering
    \includegraphics[width=\textwidth]{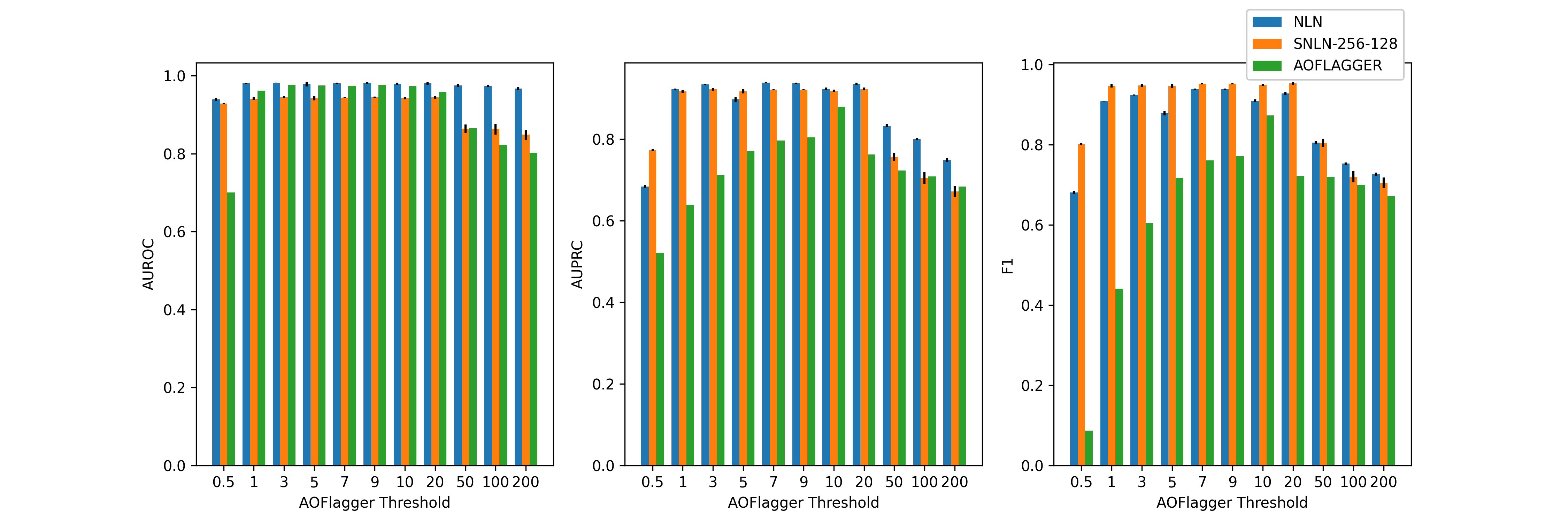}
    \caption{Comparison of NLN and SNLN methods on HERA data with all available noise sources. The AOFlagger threshold sets the baseline amplitude used to determine noise. The SNLN example used runs for 256 inference steps averaging over the last 128. Both the NLN and SNLN methods outperform AOFlagger at a low threshold, with NLN outperforming AOFlagger in all metrics at all thresholds and SNLN outperforming NLN in F1 scores in most cases and AUPRC in some cases. This demonstrates that SNLN retains the NLN's principal benefit: the ability to train and perform on over-flagged data.}
    \label{fig:threshold}
\end{figure*}
Figure \ref{fig:threshold} illustrates the performance comparison between the ANN-based NLN and our converted SNN-based SNLN approaches over a range of AOFlagger thresholds, considering all available noise sources. This threshold value affects the AOFlagger strategy to define the patches presented to the auto-encoder during training. A higher threshold will under-flag the original data, allowing RFI-contaminated patches into the auto-encoder's training data. A lower threshold will over-flag the data, limiting the variability in patches exposed to the auto-encoder. 
The results indicate that a threshold value of $10$ yields the best performance, consistent with the findings of the original work. Surprisingly, our SNLN approach consistently outperforms the NLN regarding F1 scores for all but the largest thresholds, despite the common experience that conversion from ANN to SNN may reduce accuracy \cite{bouvier_spiking_2019}. However, in our case, the binarisation of the auto-encoder output appears to enhance resilience.

\begin{table}
\caption{Performance comparison between NLN and SNLN methods on HERA data set with AOFlagger threshold of 10. Best scores in bold. The first number in SNLN entries indicates the number of inference timesteps, and the second is the number of averaged inference frames.}
\label{tab:performance:hera}
\resizebox{\textwidth}{!}{%
\begin{tabular}{cccc}
\hline
Model & AUROC & AUPRC & F1-Score \\ \hline
 AOFlagger& 0.973& 0.880&0.873\\ \midrule
NLN            & \textbf{0.983 +- 9.32e-4} & \textbf{0.940 += 6.65e-3} & 0.939 +- 1.57e-2\\ \midrule
SNLN-256-128& 0.944 +- 2.15e-3 & 0.920 +- 3.87e-3 & 0.952 +- 3.29e-3   \\ \midrule
 SNLN-256-256& 0.944 +- 2.31e-3 & 0.919 +- 4.39e-3 & 0.952 +- 3.45e-3 \\ \midrule
 SNLN-512-128& 0.944 +- 2.23e-3 & 0.920 +- 3.71e-3 & 0.952 +- 3.23e-3 \\ \midrule
 SNLN-512-256& 0.944 +- 2.27e-3 & 0.920 +- 4.02e-3 & \textbf{0.953 +- 2.87e-3} \\ \hline
\end{tabular}%
}
\end{table}
Table \ref{tab:performance:hera} presents the performance metrics of the NLN and SNLN methods on HERA data, considering all noise sources with an AOFlagger threshold of 10. The table includes AUROC, AUPRC, and F1-Score.
These results demonstrate that the SNLN conversion does not impact overall performance significantly.

\begin{figure}[htbp]
    \centering
    \includegraphics[scale=0.5]{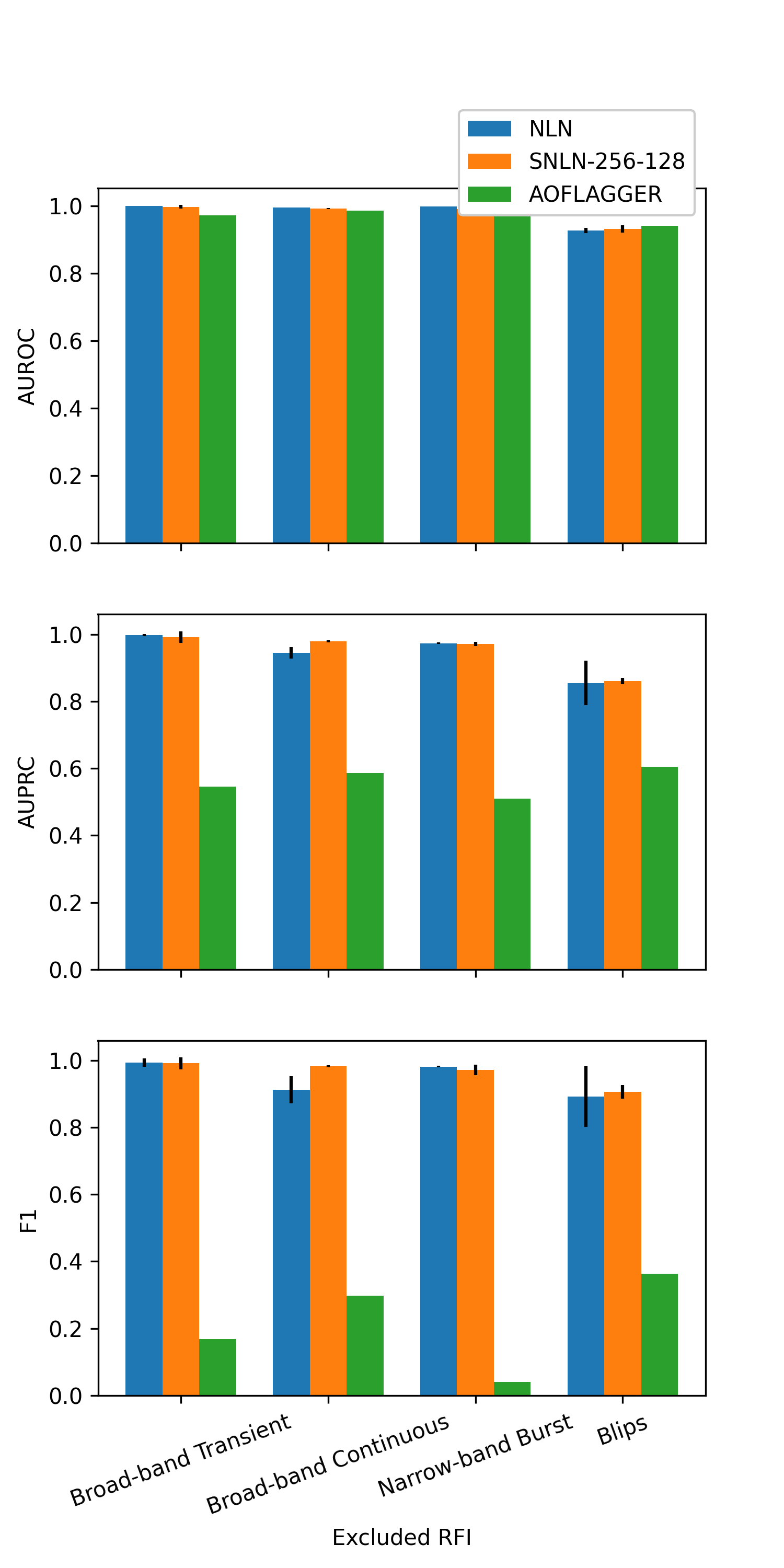}
    \caption{Out-of-distribution (OOD) performance comparison between AOFlagger and the NLN and SNLN methods. For each RFI morphology listed, all examples of that noise are withheld from the training data and exclusively present in the testing set. This test demonstrates the ability of NLN and SNLN to flag RFI that is completely unknown to the auto-encoder. The SNLN example used runs for 256 inference steps averaging over the last 128. Broad-band transient RFI is modelled on events like lighting, present across all frequencies but isolated in time. Broad-band continuous RFI is modelled on satellite communications present across a wide range of contiguous frequencies and isolated in time. Narrow-band burst RFI is modelled on ground station communication that is isolated in frequency but present over all time. Blips are isolated in frequency and time to a single impulse. See \cite{mesarcik_learning_2022} for further details on the RFI included in the HERA dataset.}
    \label{fig:ood_plot}
\end{figure}
One of the original advantages of NLN is its agnostic nature towards the type of noise present. In Figure 2 \ref{fig:ood_plot}, we showcase the AUROC, AUPRC, and F1-Scores for both NLN and SNLN methods in detecting noise from a class of RFI unseen during testing. Remarkably, our SNLN approach demonstrates strong robustness to the conversion process, producing results comparable to those reported in the original paper. Moreover, the low variance in performance observed across $10$ trials for both methods in all tests further confirms the reliability and consistency of NLN and our SNLN algorithm.
For AOFlagger, the results indicate performance for each RFI morphology in isolation. The additional skewed imbalance exacerbates any flagging weakness exhibited by the lower AUPRC and F1 scores compared to those listed in Table \ref{tab:performance:hera} and Figure \ref{fig:threshold}.

Table \ref{tab:performance:hera} also displays the variations in AUROC, AUPRC, and F1-Scores as we manipulate the inference time ($T \in \{256, 512\}$) and sample size ($k \in \{128, 256\}$).
We convert an original ANN model, trained with all noise and an AOFlagger threshold of 10 to choose uncontaminated patches in each trial. The results reveal that performance remains mostly consistent regardless of the chosen inference time and sample size beyond a specific threshold. This is consistent with the assumption that the inference time should be sufficiently long to propagate information across the spiking neurons.

\begin{figure*}[!htbp]
\centering
\begin{subfigure}{0.24\textwidth}
    \includegraphics[width=\textwidth]{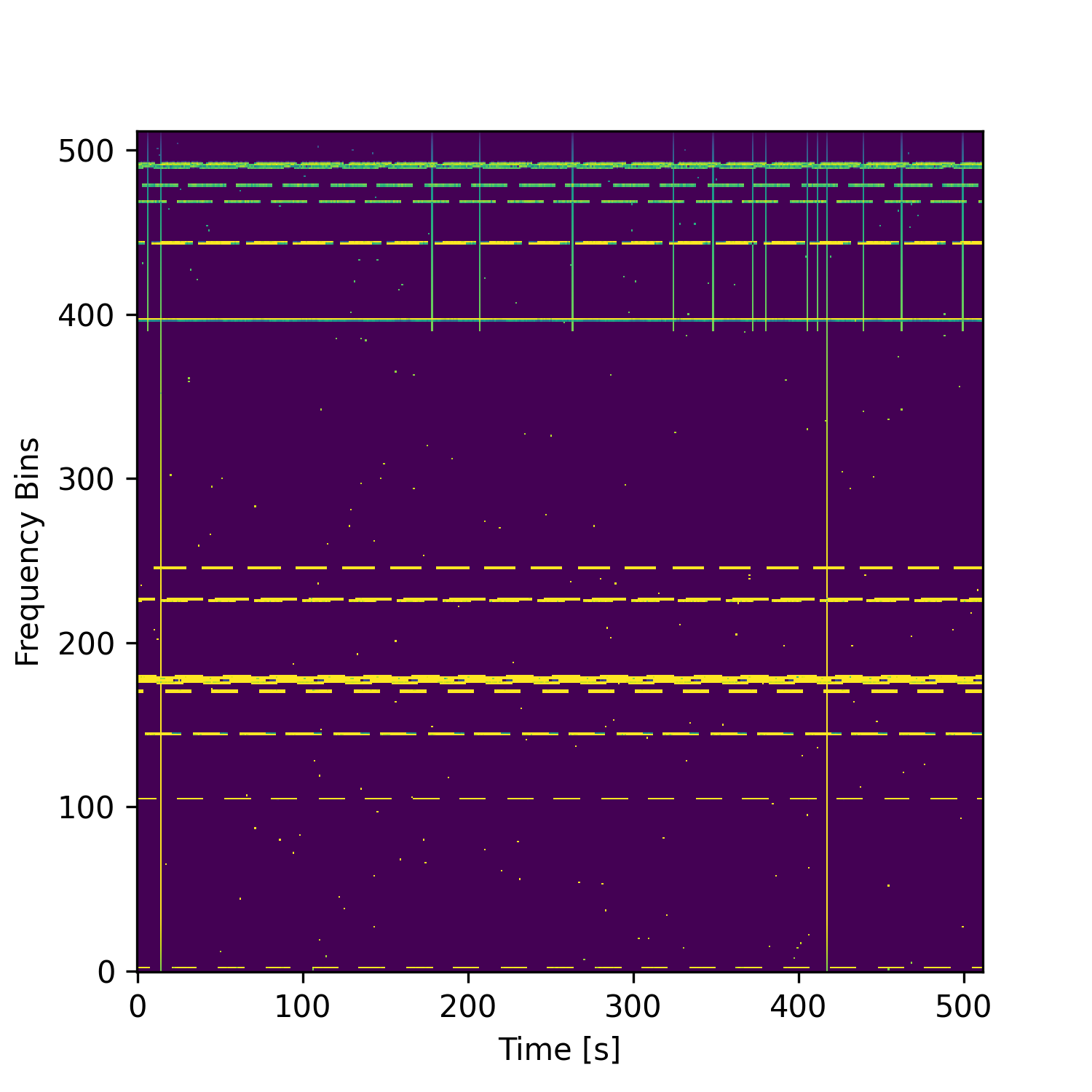}
    \caption{Original spectrum}
    \label{fig:hera:orig}
\end{subfigure}
\hfill
\begin{subfigure}{0.24\textwidth}
    \includegraphics[width=\textwidth]{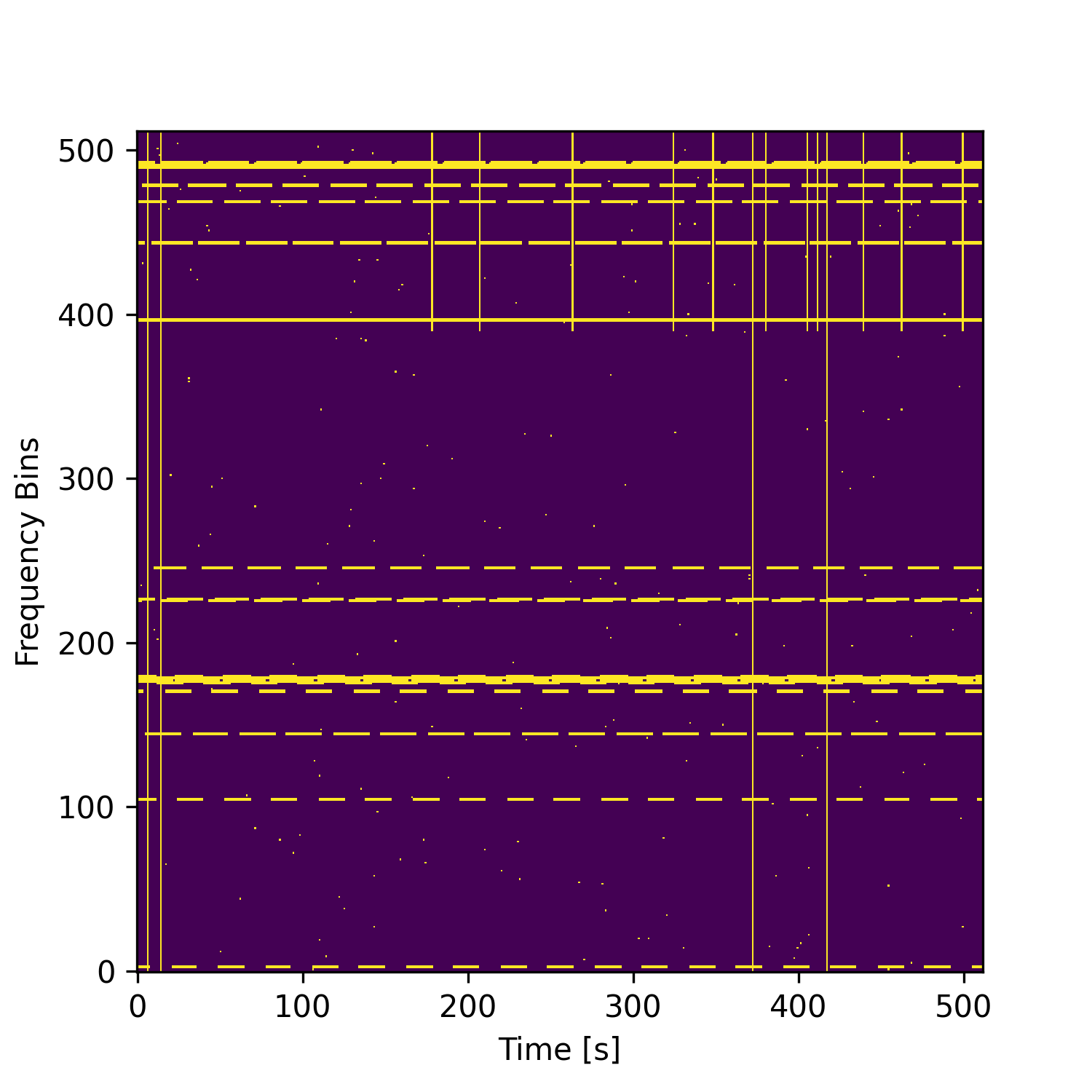}
    \caption{Ground-truth annotation}
    \label{fig:hera:mask}
\end{subfigure}
\hfill
\begin{subfigure}{0.24\textwidth}
    \includegraphics[width=\textwidth]{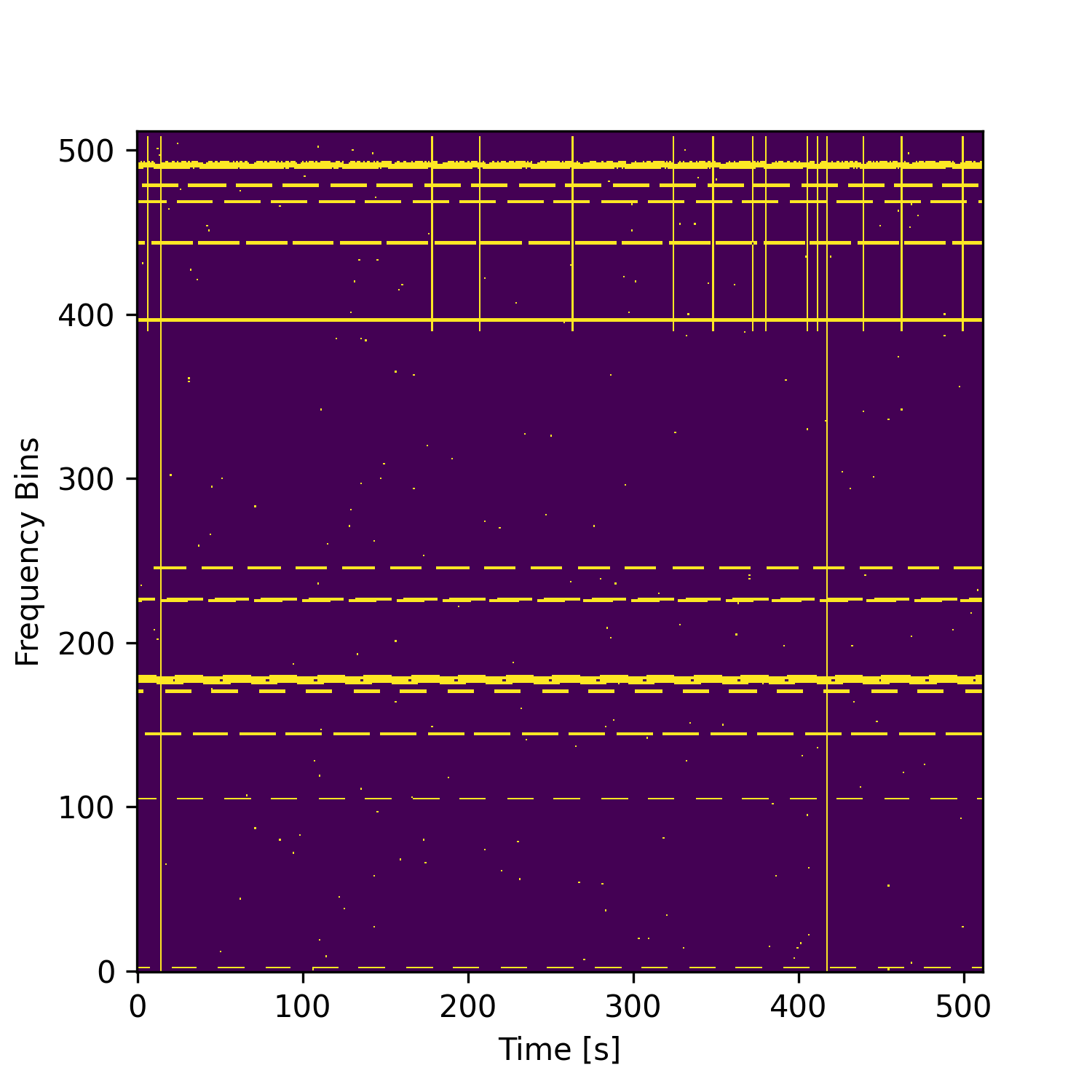}
    \caption{NLN mask}
    \label{fig:hera:nln}
\end{subfigure}
\begin{subfigure}{0.24\textwidth}
    \includegraphics[width=\textwidth]{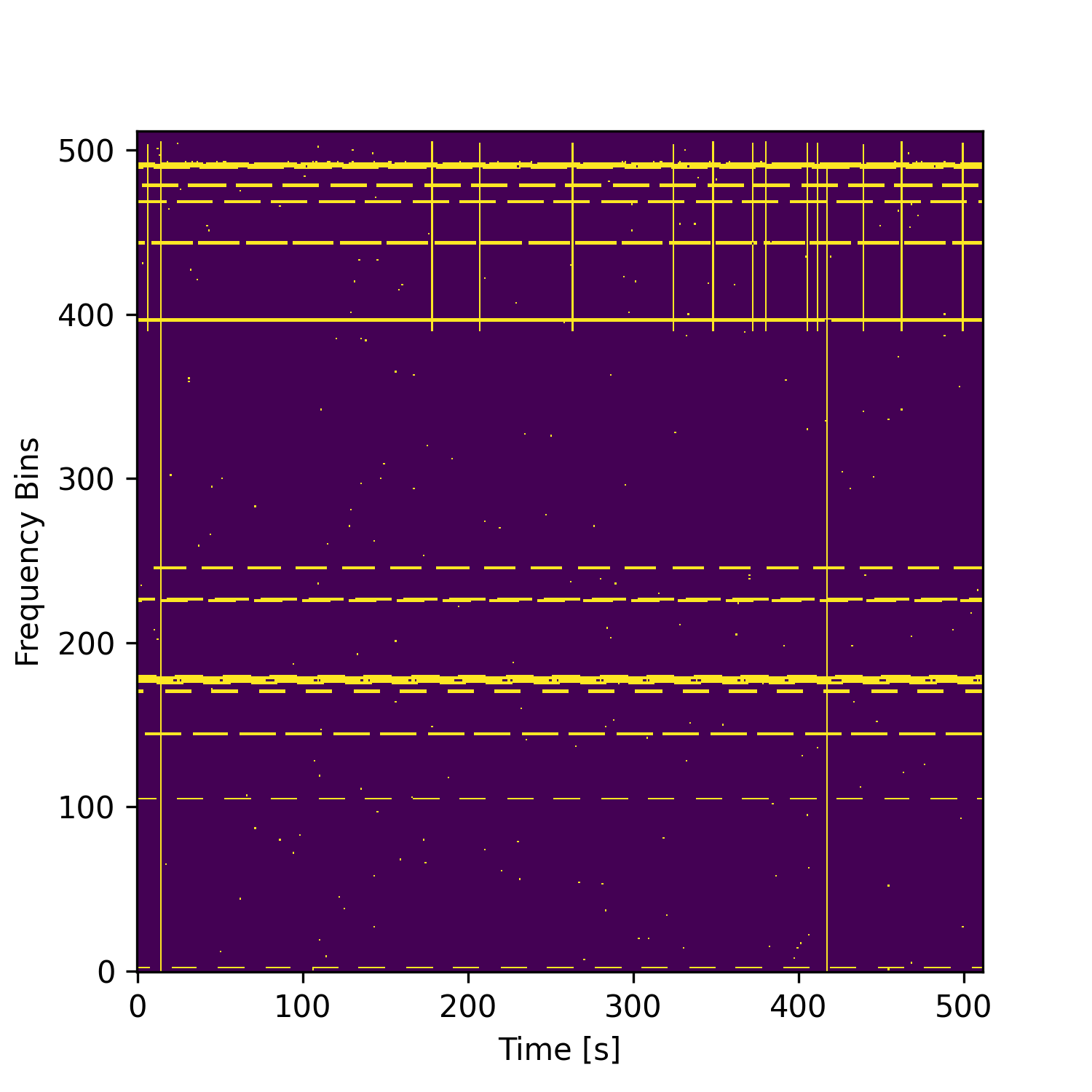}
    \caption{SNLN mask}
    \label{fig:hera:snln}
\end{subfigure}
        
\caption{Example HERA Spectrogram, the original mask, the output mask of the NLN algorithm and the output mask of the SNLN algorithm.}
\label{fig:example:hera}
\end{figure*}
Figure \ref{fig:example:hera} shows an example case from the HERA dataset. It is clear that both the NLN and SNLN methods can produce acceptable masks. 
\subsection{LOFAR}
\begin{table}
\caption{Performance comparison between NLN and SNLN methods on LOFAR data set with AOFlagger threshold of 10. Best scores in bold. The first number in SNLN entries indicates the number of inference timesteps, and the second is the number of averaged inference frames. The AOFlagger results are taken from \cite{mesarcik_learning_2022}.}
\label{tab:performance:lofar}
\resizebox{\textwidth}{!}{%
\begin{tabular}{cccc}
\hline
Model & AUROC & AUPRC & F1-Score \\ \hline
 AOFlagger& 0.788 & \textbf{0.572} & \textbf{0.570} \\ \midrule
NLN            & \textbf{0.818 +- 5.09 e-3} & 0.414 +- 2.97e-3 & 0.480 +- 6.83e-3\\ \midrule
SNLN-256-128& 0.608 +- 3.81e-2 & 0.320 +- 2.30e-2 & 0.407 +- 2.36e-2   \\ \midrule
 SNLN-256-256& 0.607 +- 3.89e-2 & 0.319 +- 2.36e-2 & 0.406 +- 2.41e-2 \\ \midrule
 SNLN-512-128& 0.609 +- 3.70e-2 & 0.321 +- 2.25e-2 & 0.408 +- 2.29e-2 \\ \midrule
 SNLN-512-256& 0.608 +- 3.73e-2 & 0.321 +- 2.25e-2 & 0.408 +- 2.30e-2 \\ \hline
\end{tabular}%
}
\end{table}
The LOFAR dataset is significantly more affected by RFI than the HERA dataset, and moreover, the RFI present is real and not simulated.
Table \ref{tab:performance:lofar} presents the performance metrics of the NLN and SNLN methods on LOFAR data, considering all noise sources with an AOFlagger threshold of 10. The table includes AUROC, AUPRC, and F1-Score.
While the NLN method outperforms AOFlagger in AUROC, we struggle to reach the same level of performance \cite{mesarcik_learning_2022} present.
Moreover, conversion from ANN2SNN incurs a more significant performance drop that persists regardless of inference and averaging time.

\begin{figure*}[!htbp]
\centering
\begin{subfigure}{0.24\textwidth}
    \includegraphics[width=\textwidth]{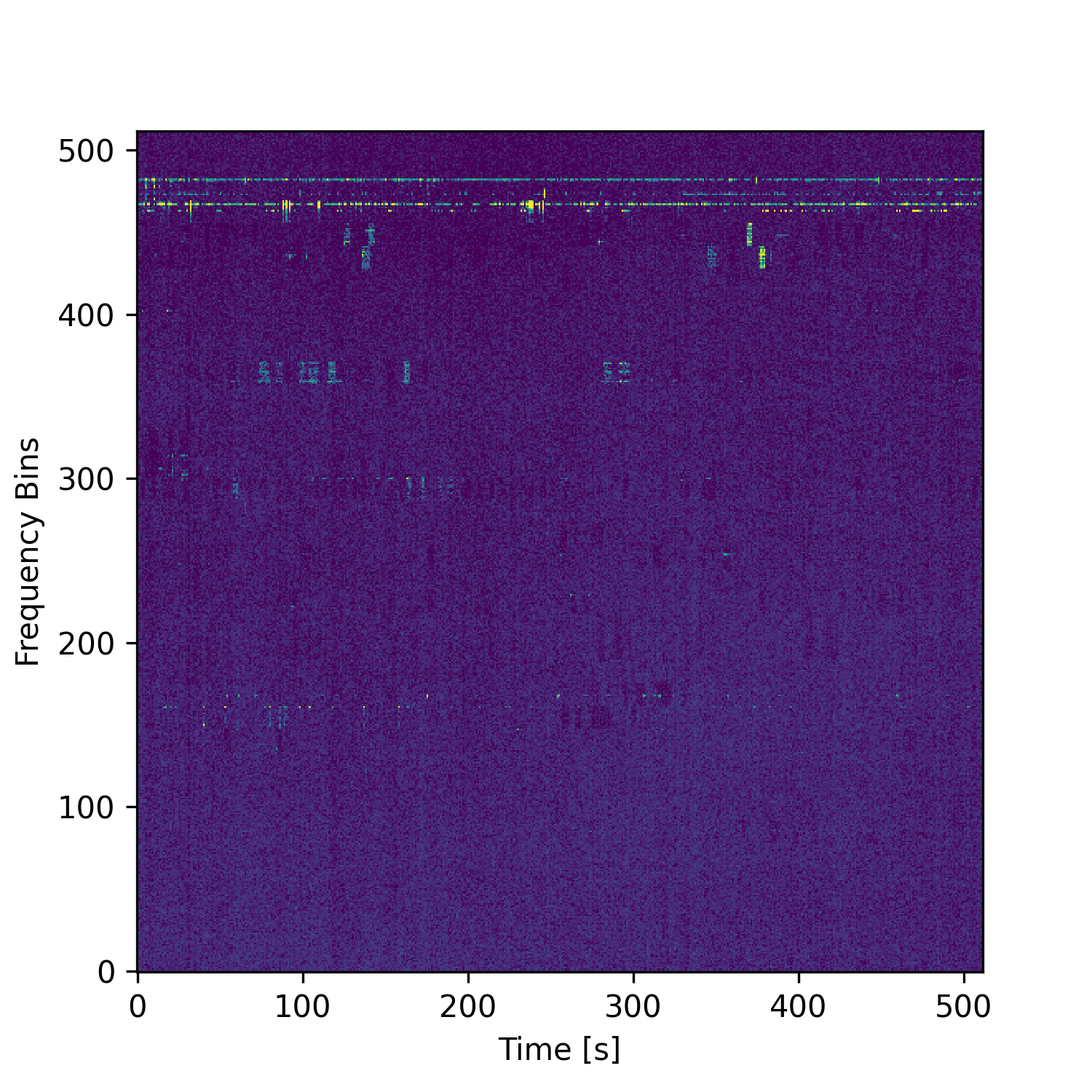}
    \caption{Original spectrum}
    \label{fig:lofar:orig}
\end{subfigure}
\hfill
\begin{subfigure}{0.24\textwidth}
    \includegraphics[width=\textwidth]{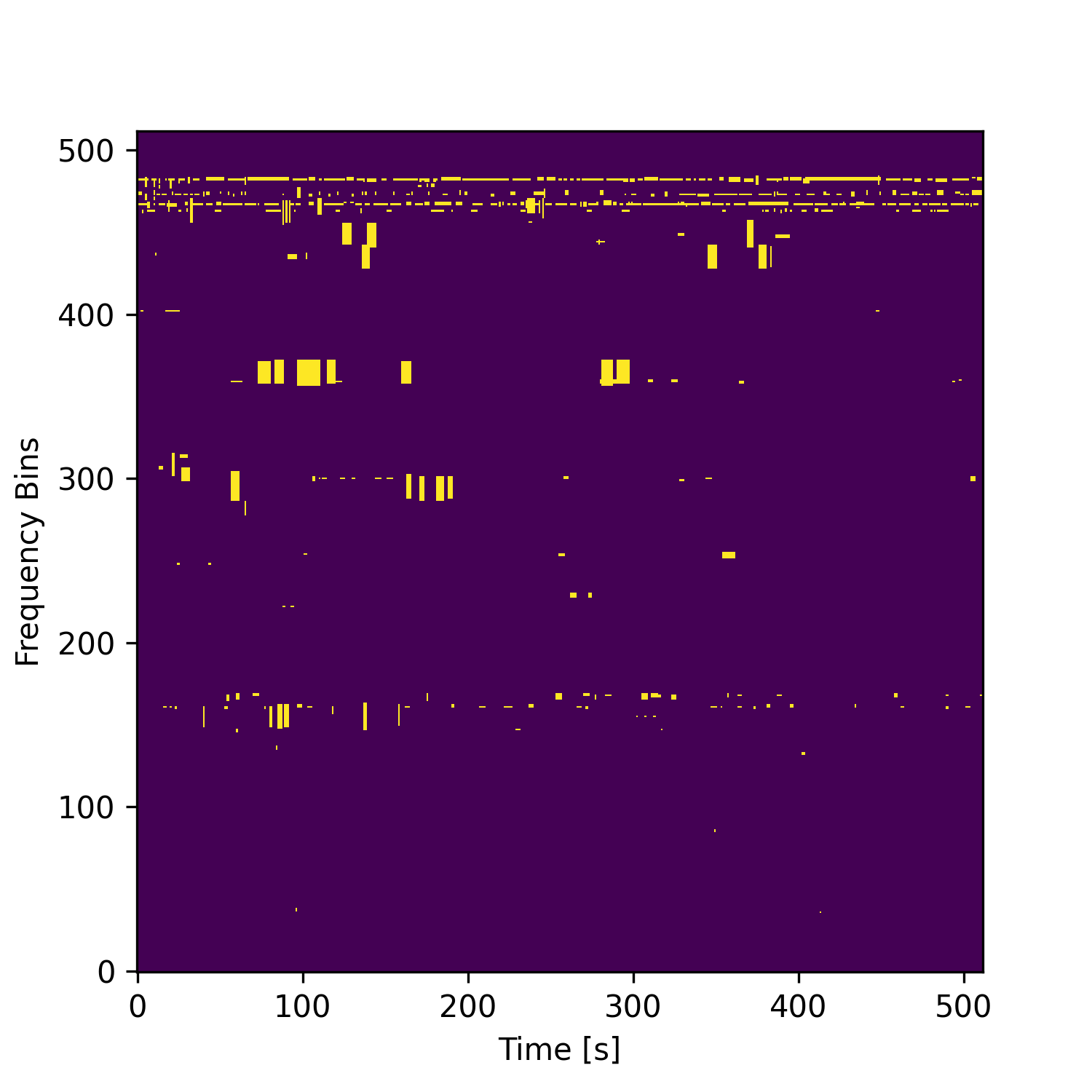}
    \caption{Expert-labelled annotation}
    \label{fig:lofar:mask}
\end{subfigure}
\hfill
\begin{subfigure}{0.24\textwidth}
    \includegraphics[width=\textwidth]{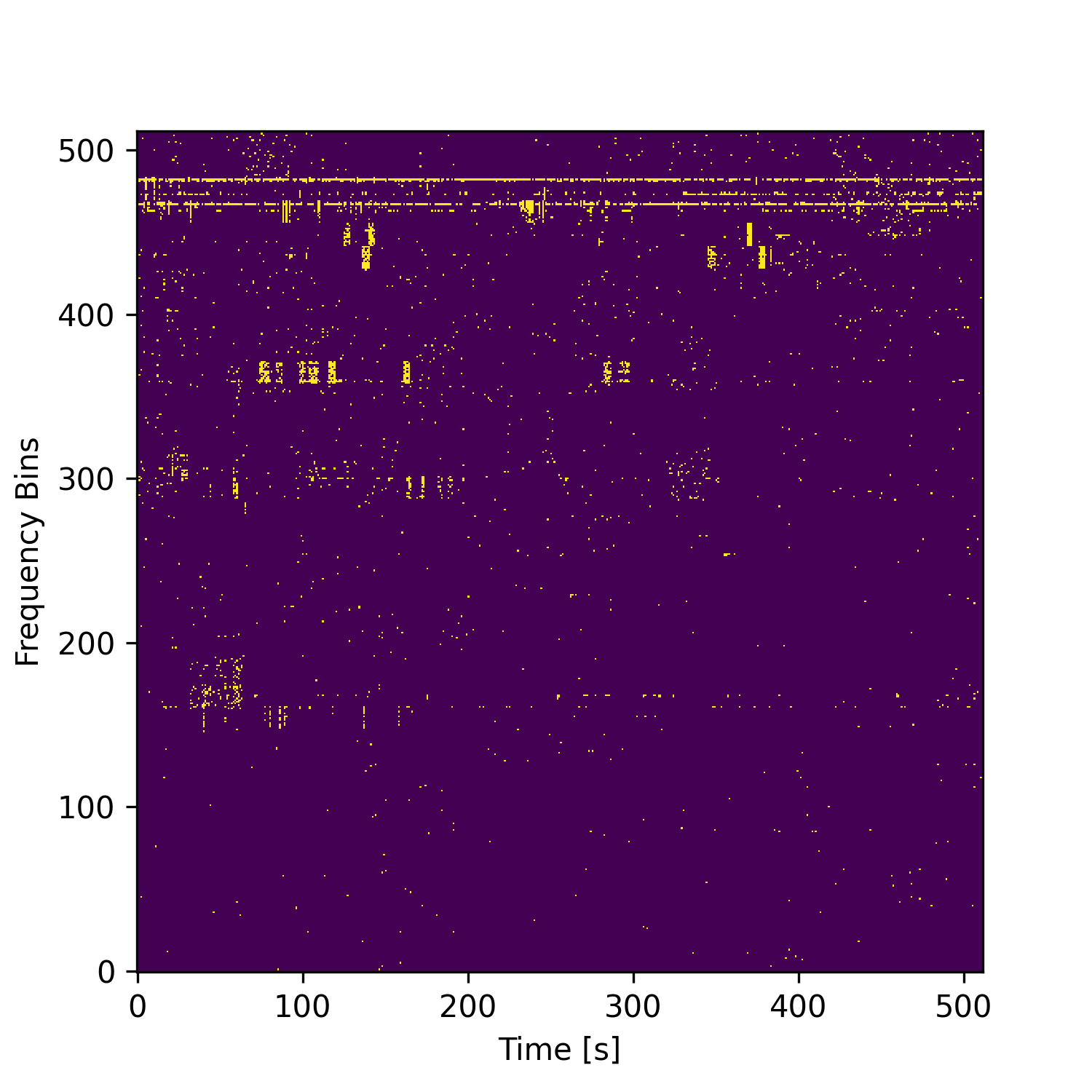}
    \caption{NLN mask}
    \label{fig:lofar:nln}
\end{subfigure}
\begin{subfigure}{0.24\textwidth}
    \includegraphics[width=\textwidth]{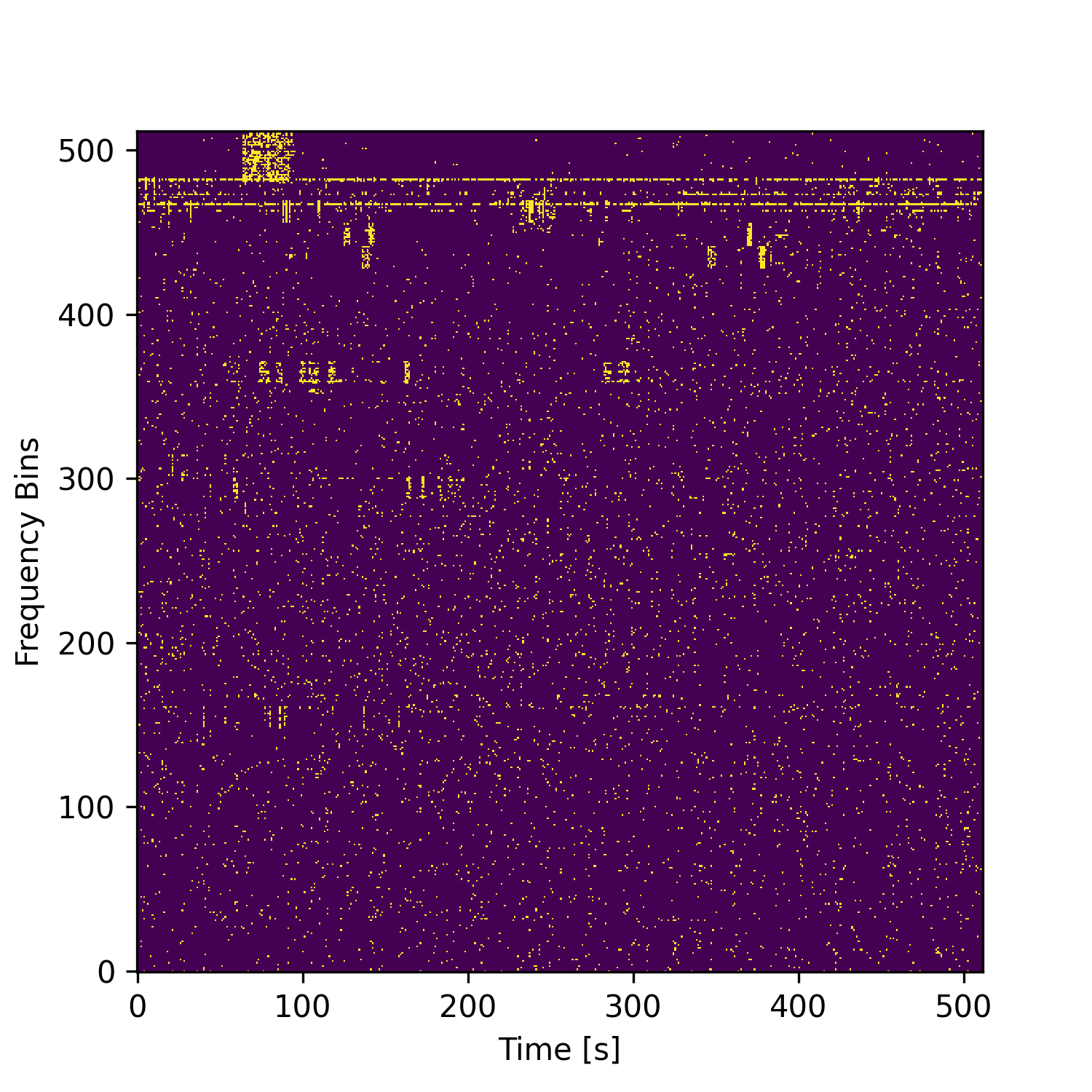}
    \caption{SNLN mask}
    \label{fig:lofar:snln}
\end{subfigure}
        
\caption{Example LOFAR Spectrogram, the original mask, the output mask of the NLN algorithm and the output mask of the SNLN algorithm.}
\label{fig:example:lofar}
\end{figure*}

Figure \ref{fig:example:lofar} shows an example case from the LOFAR dataset. The more complex background environment and the presence of more subtle RFI provide a more challenge, and here we see a discrepancy between the NLN and SNLN approaches where SNLN, on account of the multiple time-step inference, produces less sharply defined masks.

\subsection{Tabascal}
\begin{table}
\caption{Performance comparison between NLN and SNLN methods on Tabascal data set with AOFlagger threshold of 10. Best scores in bold. The first number in SNLN entries indicates the number of inference timesteps, and the second is the number of averaged inference frames.}
\label{tab:performance:tabascal}
\resizebox{\textwidth}{!}{%
\begin{tabular}{cccc}
\hline
Model & AUROC & AUPRC & F1-Score \\ \hline
 AOFlagger& 0.903 & \textbf{0.896} & \textbf{0.878} \\ \midrule
NLN            & \textbf{0.941 +- 2.20e-3} & 0.855 += 5.24e-3 & 0.830 +- 4.47e-3\\ \midrule
SNLN-256-128& 0.739 +- 2.05e-2 & 0.711 +- 1.98e-2 & 0.734 +- 1.58e-2   \\ \midrule
 SNLN-256-256& 0.743 +- 1.49e-2 & 0.712 +- 1.37e-2 & 0.735 +- 1.11e-2 \\ \midrule
 SNLN-512-128& 0.754 +- 2.45e-2 & 0.726 +- 2.19e-2 & 0.749 +- 1.64e-2 \\ \midrule
 SNLN-512-256& 0.750 +- 2.07e-2 & 0.722 +- 1.98e-2 & 0.747 +- 1.68e-2 \\ \hline
\end{tabular}%
}
\end{table}
The Tabascal dataset exhibits RFI clustered in time but over all frequencies; as such, when flagged with AOFlagger to generate training patches, the auto-encoder is never presented patches bordering on this region.
Table \ref{tab:performance:tabascal} presents the performance metrics of the NLN and SNLN methods on Tabascal data, considering all noise sources with an AOFlagger threshold of 10. The table includes AUROC, AUPRC, and F1-Score. In a case similar to that seen with LOFAR, NLN outperforms AOFlagger in AUROC but underperforms with respect to AUPRC and F1-Score. However, the discrepancy is less dramatic and the conversion to SNN causes less degradation in performance.

\begin{figure*}[!htbp]
\centering
\begin{subfigure}{0.24\textwidth}
    \includegraphics[width=\textwidth]{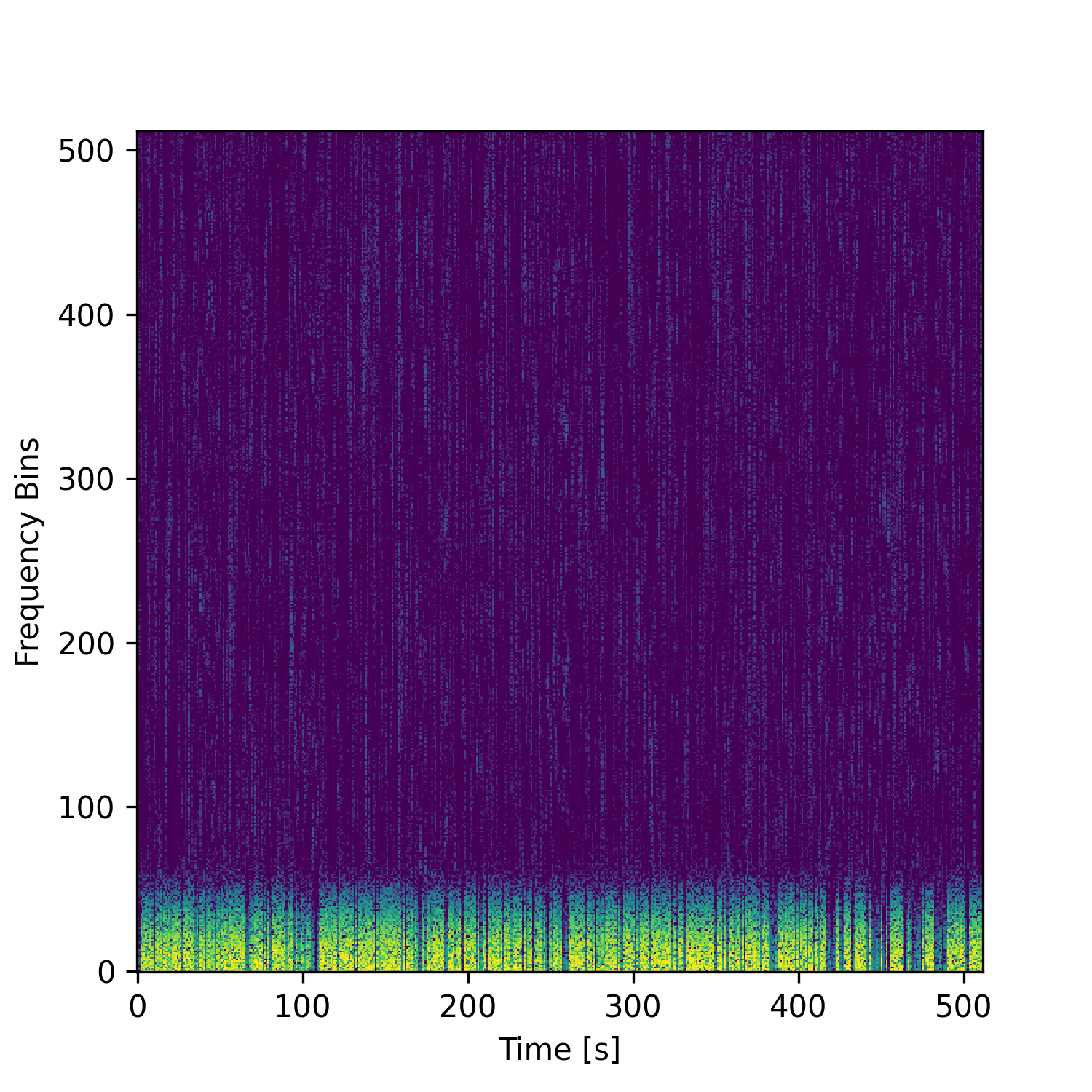}
    \caption{Original spectrum}
    \label{fig:tabascal:orig}
\end{subfigure}
\hfill
\begin{subfigure}{0.24\textwidth}
    \includegraphics[width=\textwidth]{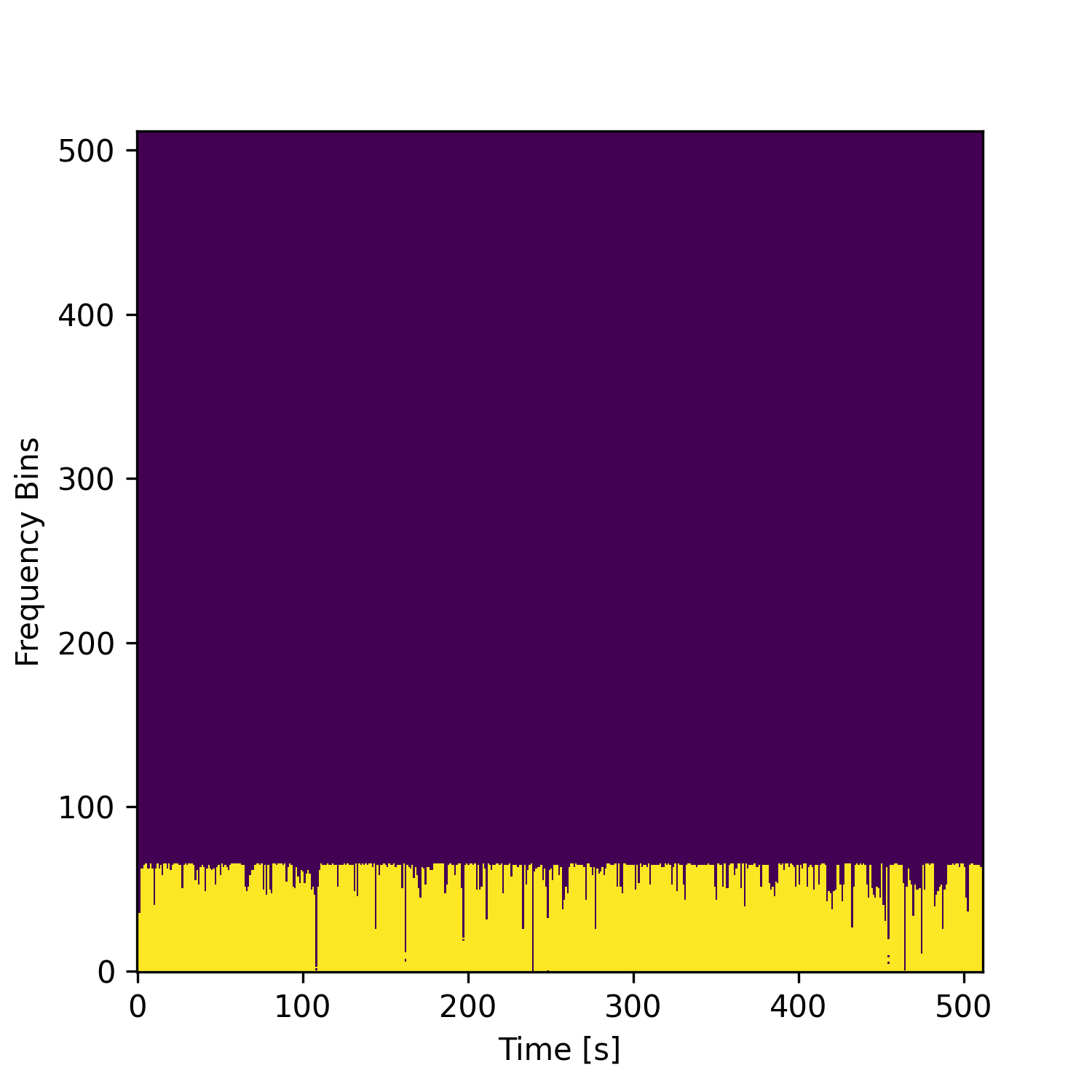}
    \caption{Ground-truth annotation}
    \label{fig:tabascal:mask}
\end{subfigure}
\hfill
\begin{subfigure}{0.24\textwidth}
    \includegraphics[width=\textwidth]{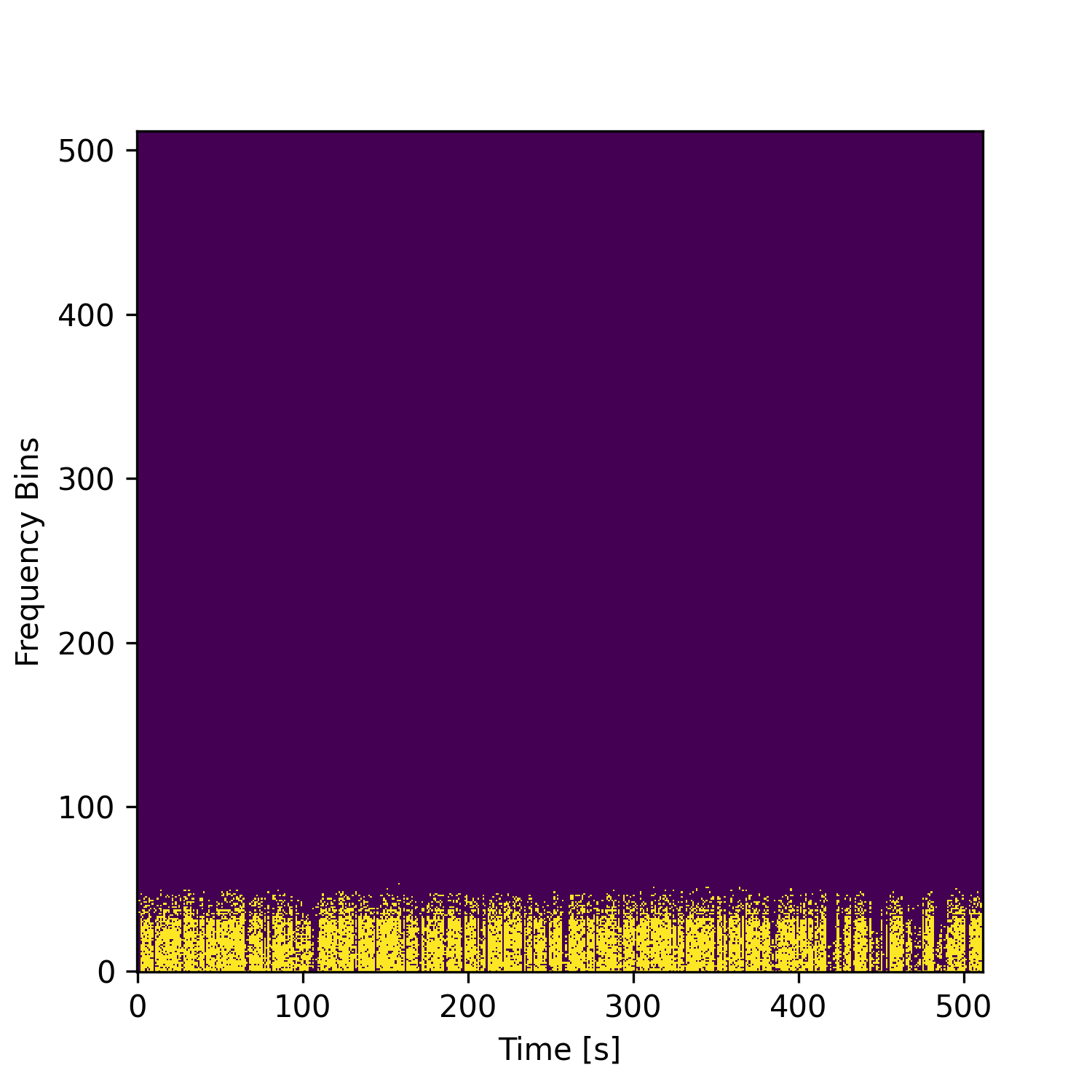}
    \caption{NLN mask}
    \label{fig:tabascal:nln}
\end{subfigure}
\begin{subfigure}{0.24\textwidth}
    \includegraphics[width=\textwidth]{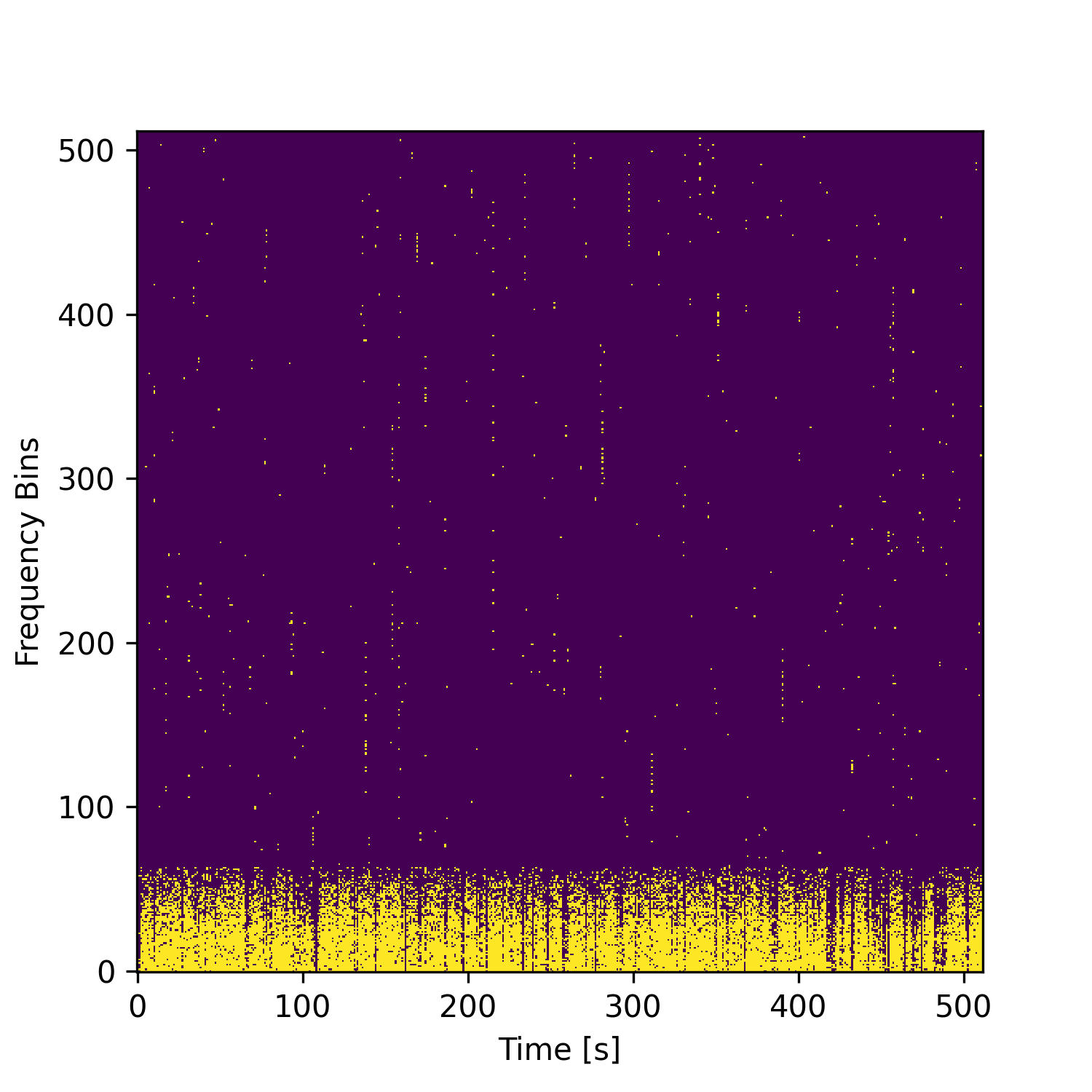}
    \caption{SNLN mask}
    \label{fig:tabascal:snln}
\end{subfigure}
        
\caption{Example Tabascal Spectrogram, the original mask, the output mask of the NLN algorithm and the output mask of the SNLN algorithm.}
\label{fig:example:tabascal}
\end{figure*}

Figure \ref{fig:example:tabascal} shows an example case from the Tabascal dataset. The RFI is localised to a specific frequency and knowledge of the precise ground-truth mask challenges the NLN and SNLN approaches. Similarly to the Lofar dataset, the SNLN produces less sharply defined masks.

\section{Conclusions}\label{sec:conclusion}
RFI detection is crucial to radio telescopes. Currently, state-of-the-art machine learning (ML) solutions treat RFI detection as a supervised semantic segmentation task. Meanwhile, real observatories utilise algorithm-based flagging solutions. \cite{mesarcik_learning_2022} proposed transforming the RFI detection problem into an anomaly detection one, introducing the NLN algorithm to address the lack of abundant high-quality labelled datasets. However, the NLN approach requires holding enough latent space samples to represent the problem domain accurately. This paper extended their approach by converting the ANN-based auto-encoder to an SNN. This conversion enables us to create a simpler downstream RFI detection scheme, SNLN, which utilises the inherently time-varying nature of SNN execution to generate latent neighbours. Our SNLN approach requires minimal data during inference time while maintaining comparable overall performance in the simpler HERA environment. Given that ANN2SNN conversion is the most straightforward way to approach using SNNs, our results provide a benchmark for SNN performance and demonstrate that SNNs offer promising potential for exploring ML-based approaches to RFI detection. It is unlikely for an SNN built by converting a trained ANN ever to exceed the original in performance. However, by providing the simplest possible demonstration of SNNs detecting RFI we lay a foundation for future work based on training fully spiking networks from scratch. Additionally, this work partially replicates the original NLN paper, providing the foundation for future benchmarks.
\subsection{Future Work}
We envisage future work to evaluate the practical performance characteristics of this SNN-based RFI detection scheme in terms of energy usage, ideally on physical neuromorphic hardware and to create new SNN-based RFI detection schemes trained from scratch. Fully spiking networks trained from scratch will exploit the time-varying nature of SNNs more and permit more sophisticated input encodings rather than being limited to rate-based encoding, as is the case when doing any ANN2SNN conversion. Moreover, this conversion introduces an additional time-axis rather than exploiting the present time-variance. Leveraging input encodings bespoke to SNNs, we believe, is key to realising the energy-efficiency and performance SNNs promise.

Moreover, in addition to methodological expansions in the application of SNNs, we anticipate future performance analysis on the RFI environments of Australian facilities such as ASKAP, the MWA and eventually SKA-Low. Such analysis should go beyond the baseline level of comparison seen in this article and move towards a more robust comparison and potential fusion with operational RFI flagging schemes.

Finally, to our knowledge, this work marks the first application of SNNs to any data-processing or scientific task in Astronomy; we hope this technology finds use in more parts of the spatiotemporal world, as seen in the radio sky.

\paragraph{Funding Statement}
This work was supported by a Westpac Future Leaders Scholarship, an Australian Government Research Training Program Fees Offset and an Australian Government Research Training Program Stipend.
\paragraph{Competing Interests}
None
\paragraph{Data Availability Statement}
The current codebase is available on Git Hub\footnote{\url{https://github.com/pritchardn/SNN-NLN}}.
The code used in this publication to generate the Tabascal dataset is available on Git Hub\footnote{\url{https://github.com/pritchardn/tabascal-dataset}}.
The Tabascal dataset is available on Zenodo\footnote{\url{https://doi.org/10.5281/zenodo.8401763}}, as are all results\footnote{\url{https://doi.org/10.5281/zenodo.10060036}}.

\printendnotes

\bibliography{main}

\appendix

\end{document}